\documentclass[twocolumn,aps,pra,superscriptaddress,showpacs,tightenlines]{revtex4}
\usepackage{amsmath}
\usepackage{amsfonts}
\usepackage{graphicx}
\usepackage{epsfig}
\usepackage{color}
\usepackage[colorlinks,citecolor=blue]{hyperref}

\begin{document}

\title{Superfluid-Mott-insulator transition in superconducting circuits with weak anharmonicity}

\author{Li-Li Zheng}
\affiliation{School of physics, Huazhong University of Science and Technology, Wuhan 430074, China}

\author{Ke-Min Li}
\affiliation{School of physics, Huazhong University of Science and Technology, Wuhan 430074, China}
\affiliation{Department of Physics, Zhejiang University, Hangzhou 310027, China}

\author{Xin-You L\"{u}}
\email{xinyoulu@hust.edu.cn}
\affiliation{School of physics, Huazhong University of Science and Technology, Wuhan 430074, China}

\author{Ying Wu}
\email{yingwu2@126.com}
\affiliation{School of physics, Huazhong University of Science and Technology, Wuhan 430074, China}

\date{\today}

\begin{abstract}
We investigate theoretically the ground-state property of a two-dimensional array of superconducting circuits including the on-site superconducting qubits (SQs) with weak anharmonicity.
In particular, we analyse the influence of this anharmonicity on the Mott insulator to superfluid quantum phase transition.
The complete ground-state phase diagrams are presented under the mean field approximation. Interestingly, the anharmonicity of SQs affects the Mott lobes enormously, and the single excitation Mott lobe disappears when the anharmonicity become zero. Our results can be used to guide the implementations of quantum simulations using the superconducting circuits, which have nice integrating and flexibility.
\end{abstract}
\maketitle

\section{introduction}
Quantum phase transition (QPT), occurring at nearly zero temperature, plays an important role in many areas of physics, including particle physics, condense matter, and quantum optics. It has been extensively studied in the interacting systems, such as heavy fermions in Kondo lattices~\cite{Rosch2007}, ultracold atoms in optical lattices~\cite{Altman2003,Han2004,Orth2008}, and the ensemble of two-level systems interacting with a bosonic field (i.e., Dicke model)~\cite{Dicke1954}. In particular, the phase transition between Mott insulator phase and superfluid phase is predicted in the Bose-Hubbard model~\cite{Fisher1989}. It originally comes from the competition between the Kerr-nonlinearity-induced on-site photon-photon repulsion (i.e., photon blockade~\cite{Schmidt1997}) and photon hopping effects between the neighboring sites. Beside the Kerr nonlinearity medium, the Jaynes-Cummings (JC) model, describing the interaction between a single-mode bosonic field with a two-level atom/qubit, also could offer the photon-photon repulsion~\cite{Tan2002}, and the photon blockade in JC model has been experimentally realized~\cite{Birnbaum2005}. Then the JC-Hubbard model is proposed for demonstrating the Mott insulator to superfluid quantum phase transition~\cite{Greentree2006,Na2008,Shen2016}. The similarities and differences between the JC-Hubbard model and the Bose-Hubbard model also have been discussed in detail~\cite{Koch2009}.
\begin{figure}[htb]
\centering\includegraphics[width=8cm]{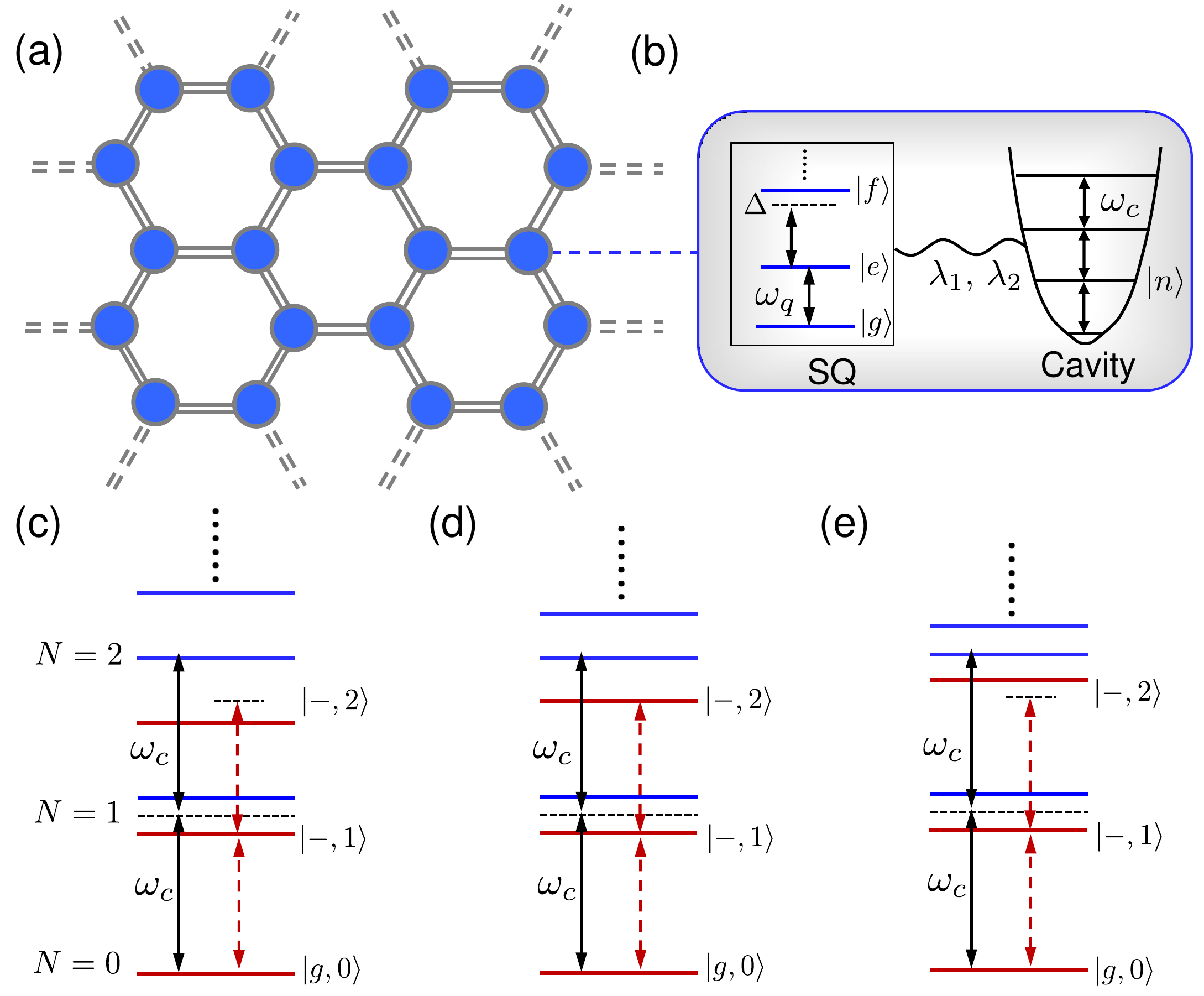}
\caption{(color online) (a) Schematic diagram of a two-dimensinal lattice, consisting of an array of microwave cavities with the nearest-neighbor photon hopping. (b) Each cavity (with frequency $\omega_c$) interacts a SQ (with frequency $\omega_q$) with coupling strengths $\lambda_1$ and $\lambda_2$. Here the SQ has weak anharmonicity $\Delta=\omega_{ef}-\omega_{ge}$, and then the third level $|f\rangle$ is considered. Also, we consider the high levels anharmonicity are much larger than $\Delta$ and then we only consider three levels $|g\rangle$, $|e\rangle$ and $|f\rangle$. The coupling strengths $\lambda_1$ and $\lambda_2$ correspond to the qubit-cavity coupling strengths between transitions $|g\rangle\rightarrow|e\rangle$, $|e\rangle\rightarrow|f\rangle$ and cavity mode, respectively. (c-e) The energy structure of on-site excitations on the resonant condition $\delta=\omega_q-\omega_c=0$ and (c) $\Delta<0$, (d) $\Delta=0$, (e) $\Delta>0$.}
\label{Fig1}
\end{figure}

Generally, the quantum criticality in the strongly correlated many-particle system is very hard to be demonstrated experimentally. Quantum simulation employing a controlled quantum mechanical device can mimic and investigate the quantum property of other systems~\cite{Buluta2009}, which offers a method to explore the quantum criticality of the strongly correlated system. Superconducting circuits based on Josephson junctions are promising candidates for the implementation of JC-Hubbard model, owing to their large-scale integration, design flexibility and easy manipulation~\cite{You2005,Makhlin2001,Clarke2008,Schoelkopf2008,You2011,Ladd2010}. The quantum phase transition property has been studied in the coupled superconducting circuit lattices including the on-site JC interaction between the SQs and the microwave resonator~\cite{Houck2012,Jin2013,Yang2014,Deng2016,Seo2015}. However, many SQs with nice coherent time (e.g., phase qubits~\cite{Martinis2002,Joo2010}, capacitively shunted flux qubits~\cite{You2007,Ste2010} and transmon qubits~\cite{Koch2007}) have a weakly anharmonic energy-level-structure (i.e., the detuning between adjacent transition frequencies is very small), and the two-level approximation is invalid. Recently, there have been a number of theoretical studies analyzing the effects of the weak anharmonicity of SQs on the quantum gate operations~\cite{Fazio1999,Steffen2003,Zhou2005,Rebentrost2009,Ferr2010,Ashhab2012}. Similarly, the weak anharmonicity of SQs will also influence the ground-state property of the coupled circuit lattices of implementing the JC-Hubbard model.
\begin{figure*}[htb]
\includegraphics[scale=0.35]{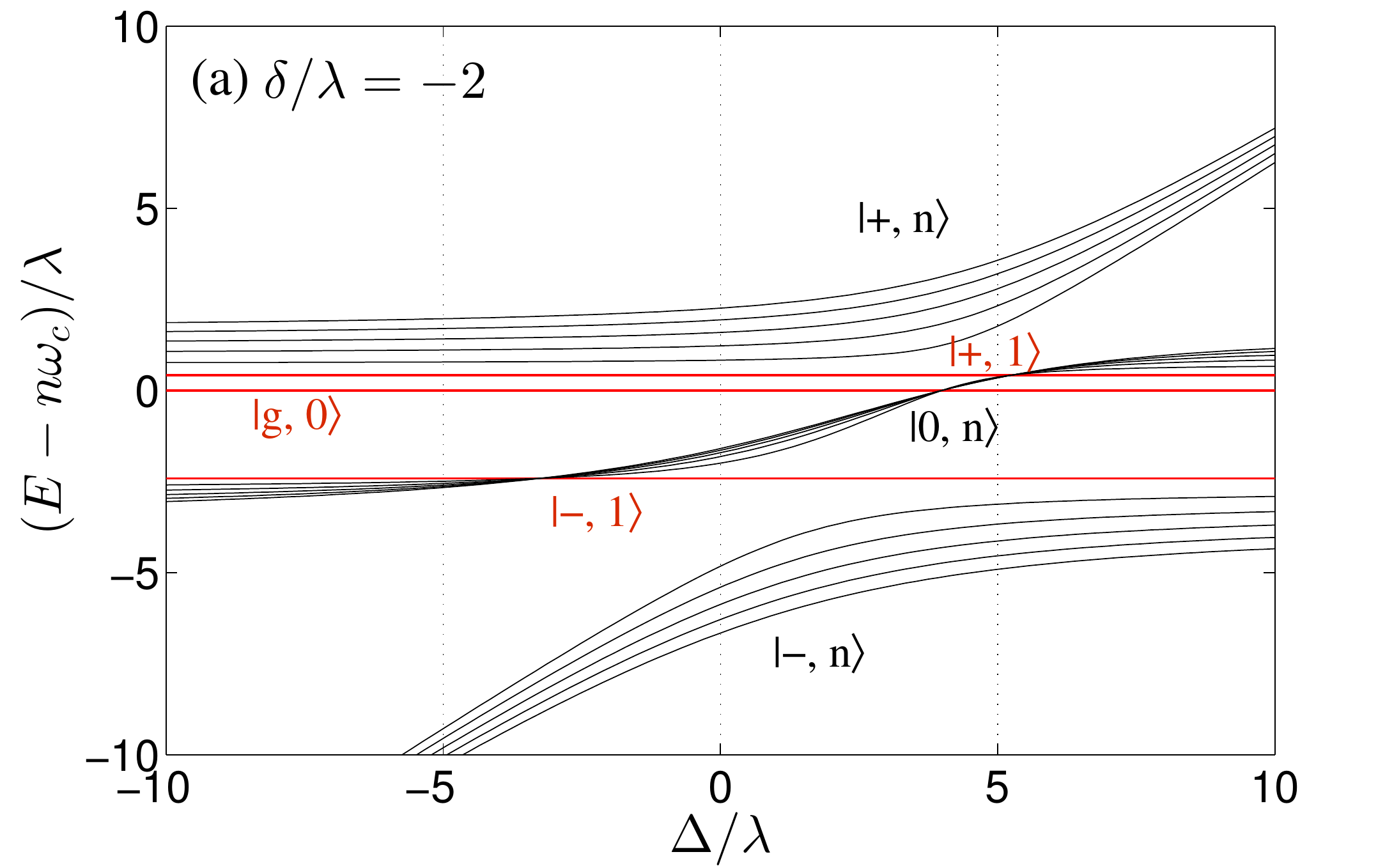}\includegraphics[scale=0.35]{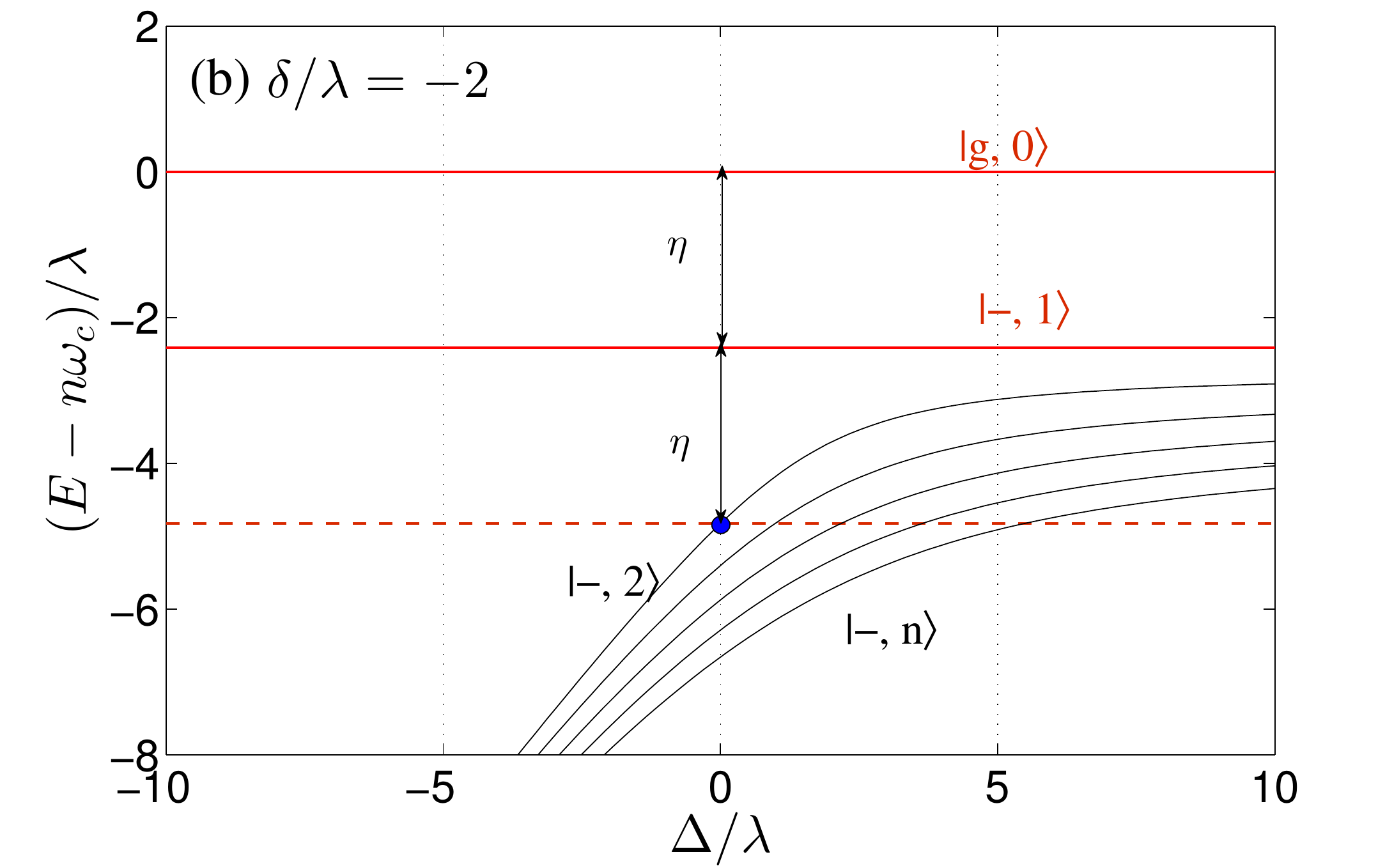}
\includegraphics[scale=0.35]{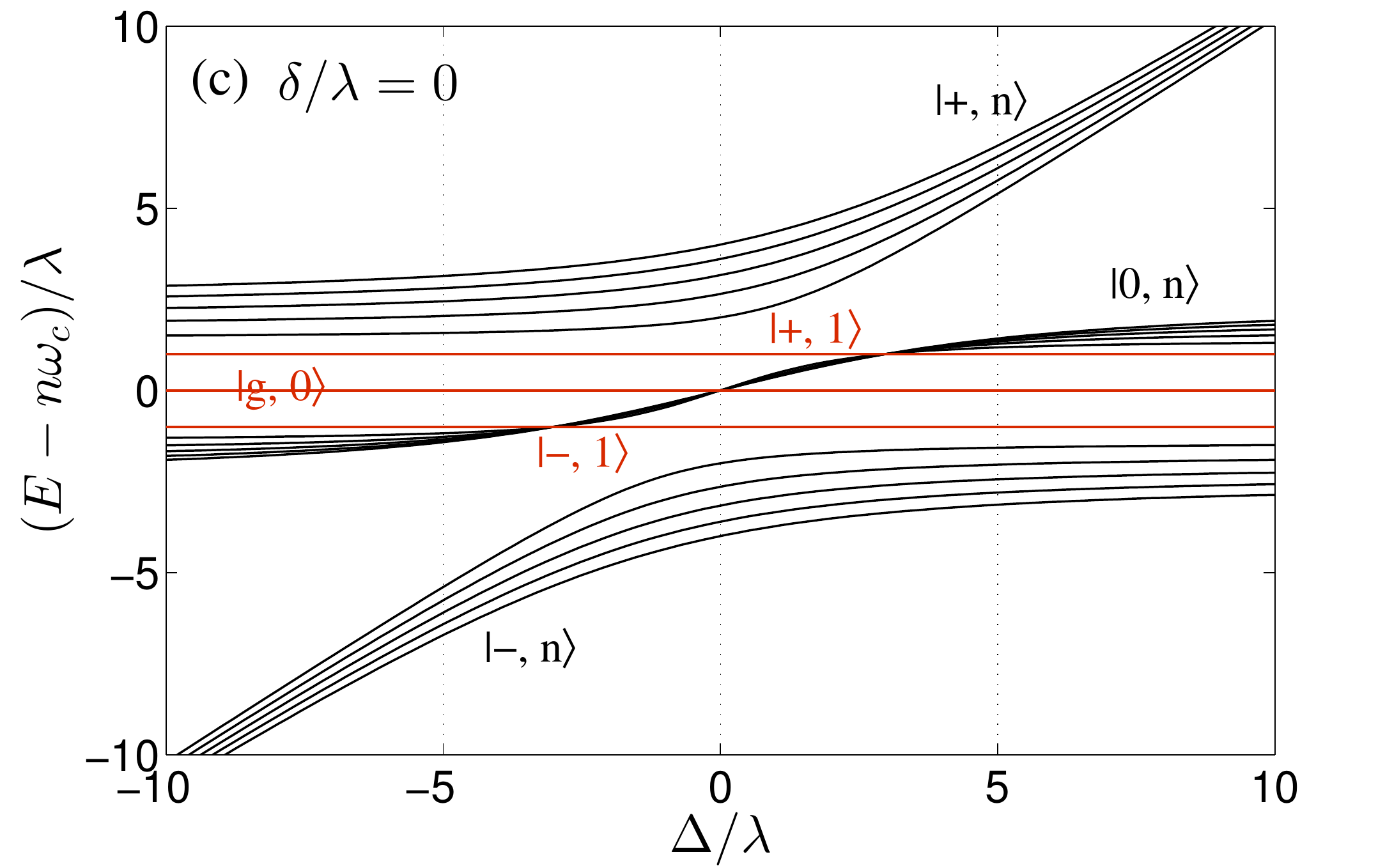}\includegraphics[scale=0.35]{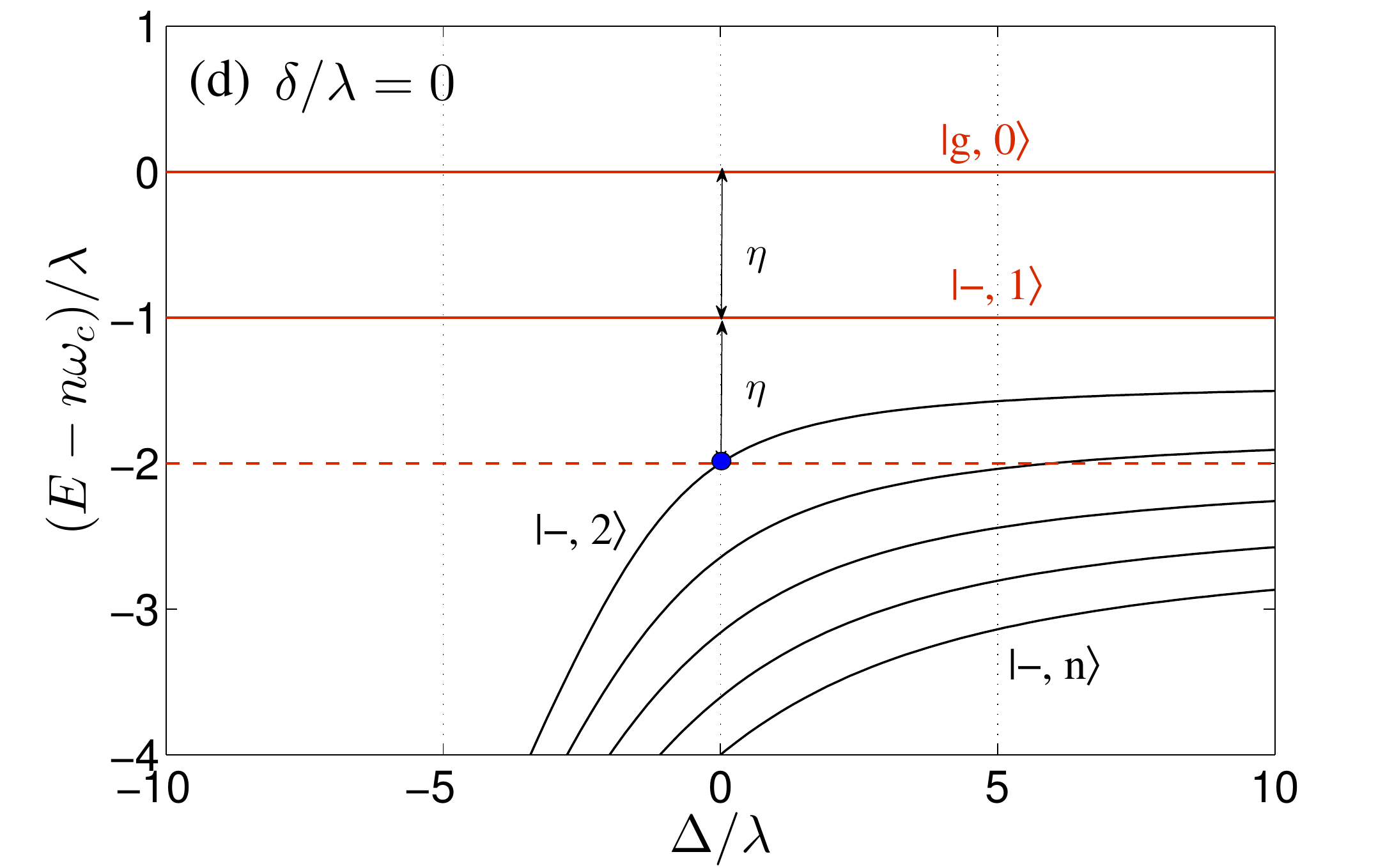}
\includegraphics[scale=0.35]{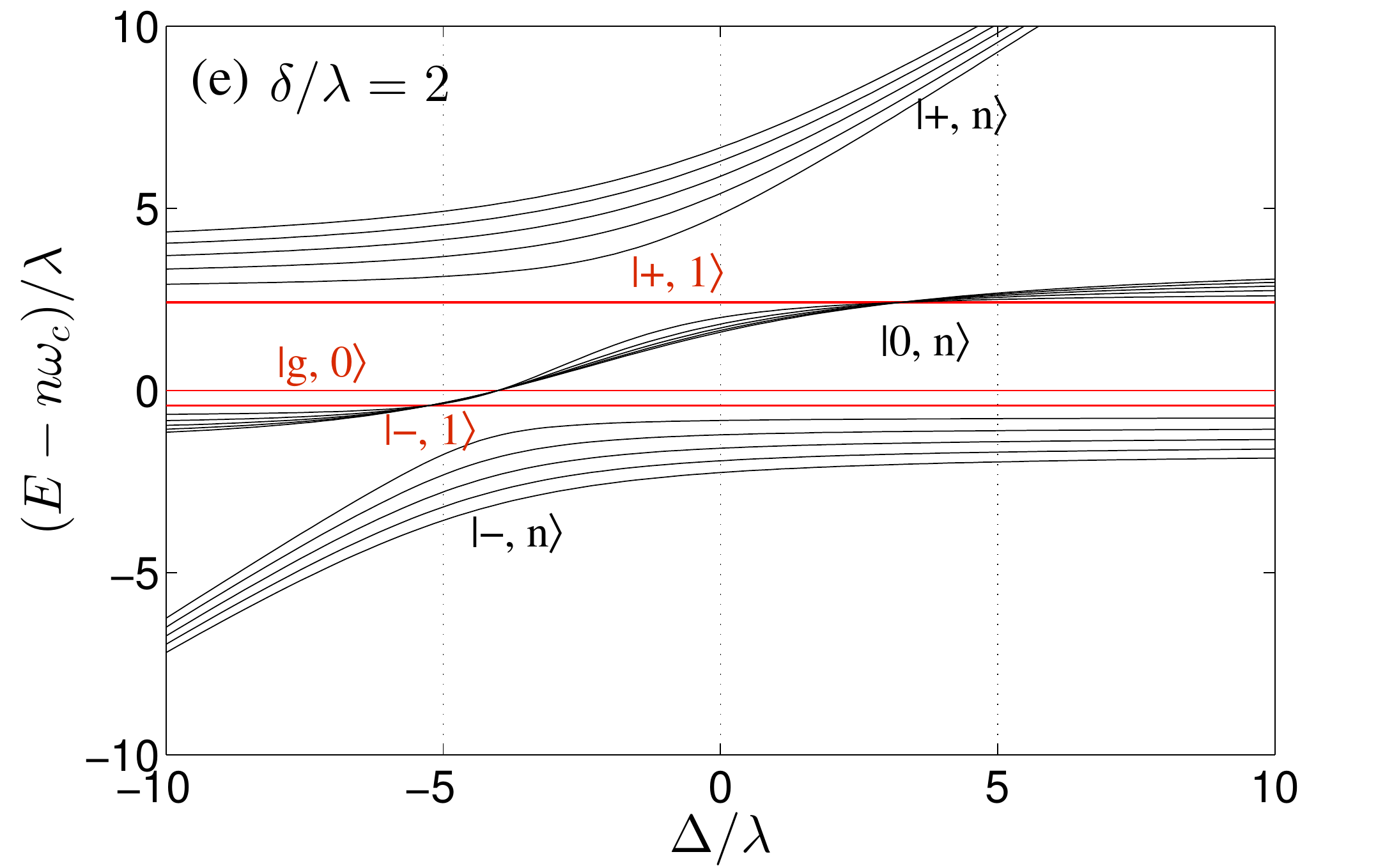}\includegraphics[scale=0.35]{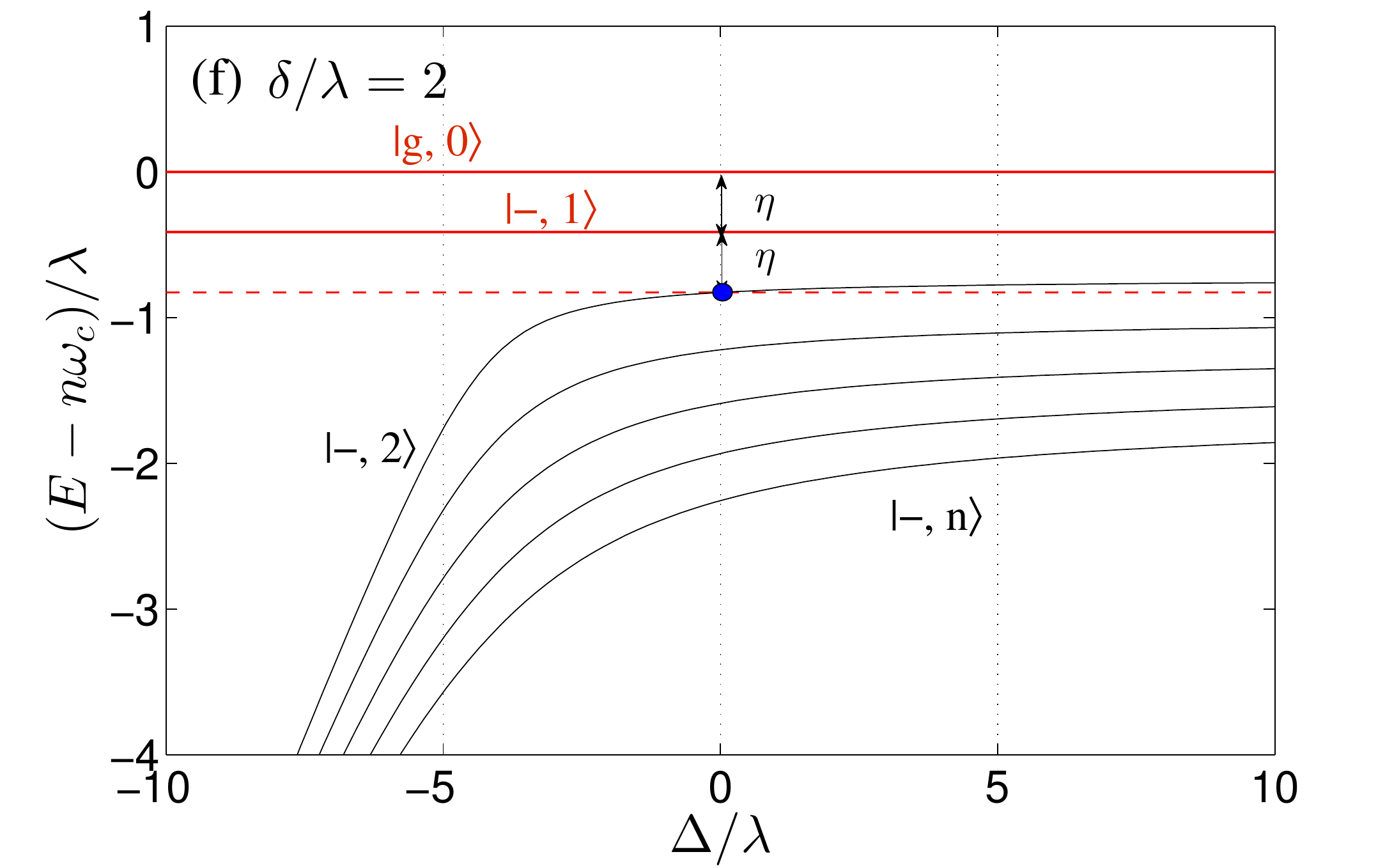}
\caption{The eigenvalues of system as a function of anharmonicity of SQ (i.e., $\Delta$), when the qubit-resonator detuning (a,b) $\delta=-2$, (c,d) $\delta=0$, and (e,f) $\delta=2$.
It splits into three branches denoted by $|+,n\rangle$ (the upper branch), $|0,n\rangle$ (the middle branch), and $|-,n\rangle$ (the lower branch).
Note that, the eigenvalues corresponding to the single excitation subspace are flat with respective to $\Delta$.
The subplots (b,d,f) indicate the corresponding ground states of system corresponding to the fixed excitation number $N_i$. The single excitation nonlinearity is denoted by $\eta$.}
\label{Fig2}
\end{figure*}

\begin{figure}[htb]
\includegraphics[scale=0.55]{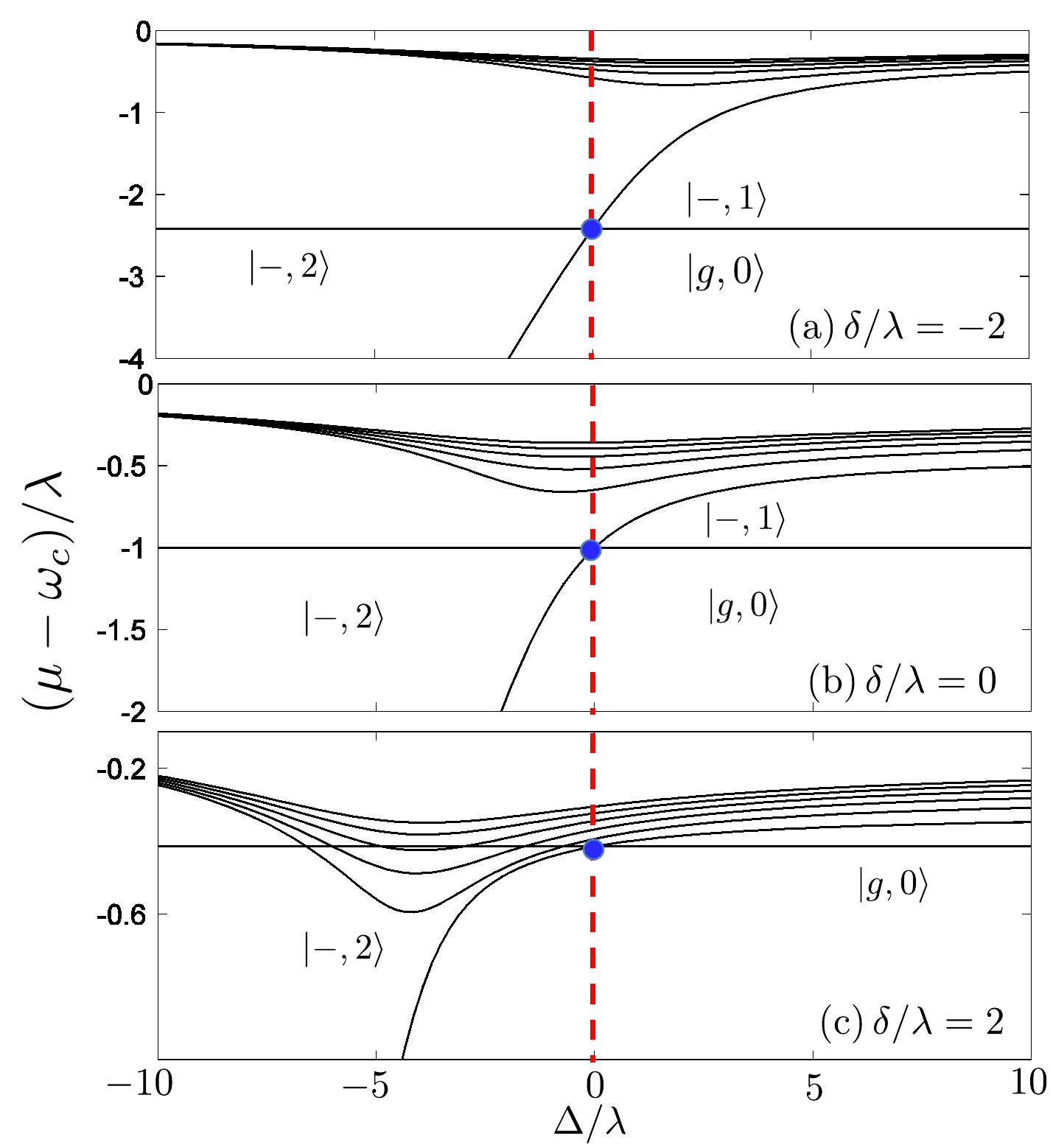}
\caption{Boundaries between Mott lobes as a function of anharmonicity $\Delta$ and chemical potential $\mu$ in the limit of small $J$.
The crossing denoted by blue dotes indicate the position where the single excitation Mott lobes disappears.
Before and after this crossing position the lowest stable regions correspond to $|-,2\rangle$ and $|g,0\rangle$, respectively.}
\label{Fig3}
\label{boundary}
\end{figure}

\begin{figure}[htb]
\includegraphics[scale=0.55]{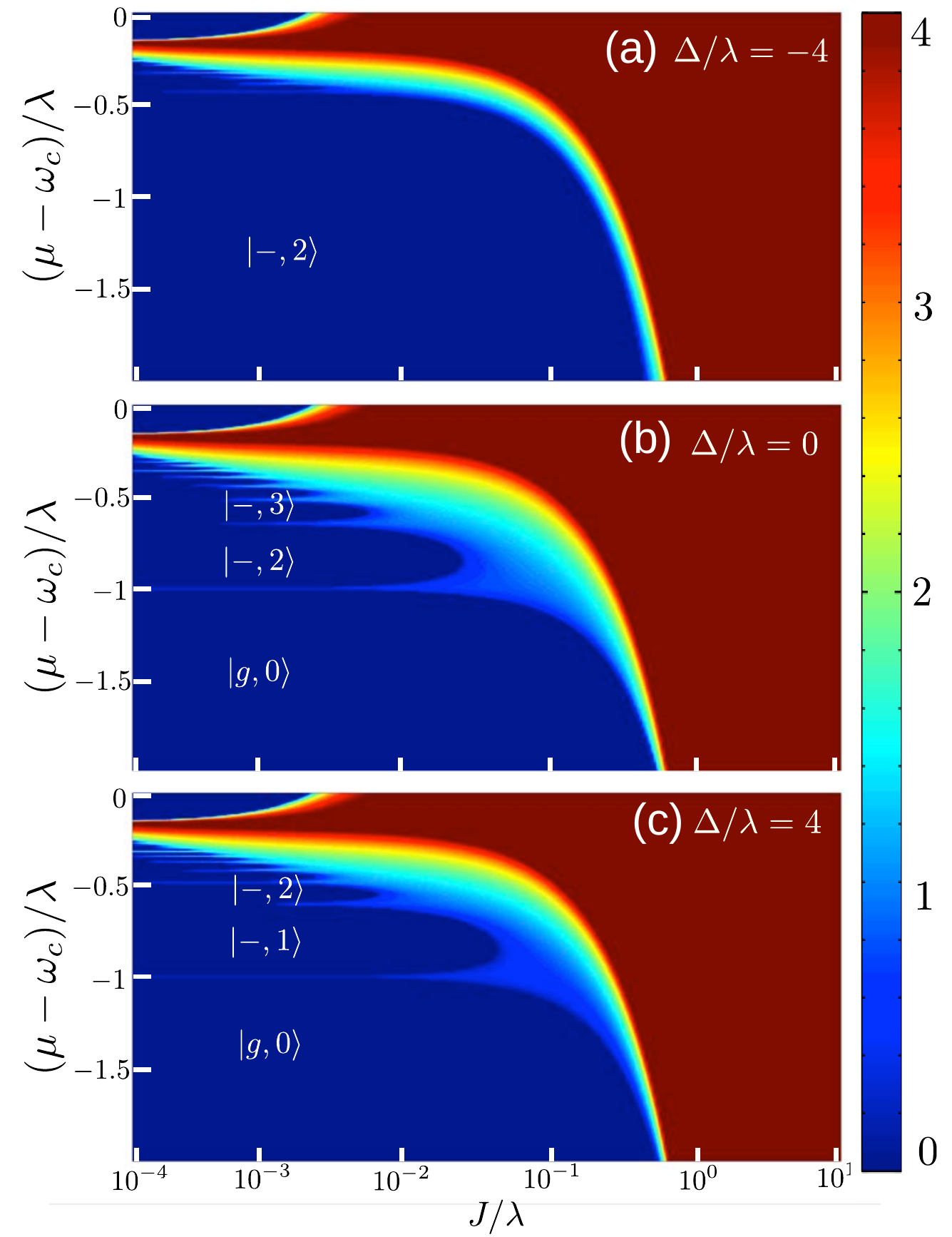}
\caption{The superfluid order parameter $\psi$ as a function of the photon hopping $J$ and the chemical potential $\mu$ for different $\Delta$.
The Mott-insulator lobes and superfluid phases are denoted by the regions of $\psi=0$ and $\psi\neq0$, respectively.
In the left-hand edge the system is in the Mott-insulator phase due to the photonic repulsion dominates over hopping.
The superfluid phase is on the right-hand edge. Here we have chosen the qubit-cavity resonance $\delta=0$, and then this phase diagram corresponds to the regime of Fig.\,\ref{Fig3}(b).}
\label{Fig4}
\end{figure}

\begin{figure*}
\includegraphics[scale=0.38]{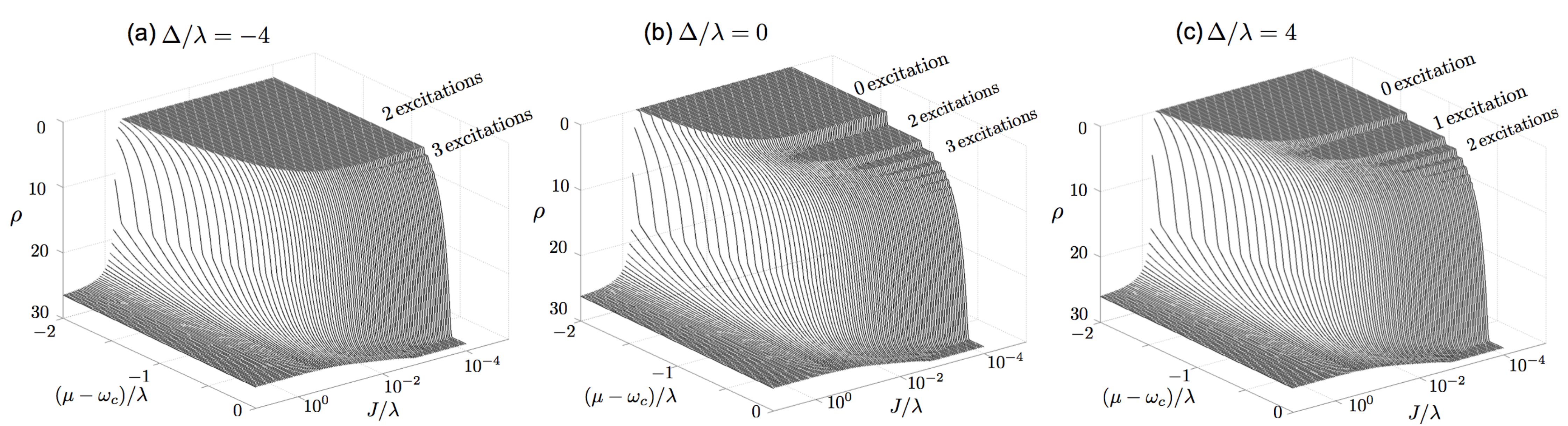}
\caption{The on-site excitations of Mott lobes $\rho$ as a function of photon hopping rate $J$ and chemical potential $\mu$ for different $\Delta$.
The plateaux with constant $\rho$ indicate the regions of Mott lobes with fixed on-site excitation number.
In the case of $\Delta=0$, the first, second and third stair steps correspond to the states of Mott lobes are $|g,0\rangle$, $|-,2\rangle$, and $|-,3\rangle$ for 0, 2, 3 excitations, respectively.
Corresponding to Fig.\,\ref{Fig4}, the resonant condition $\delta=0$ has been chosen here.}
\label{Fig5}
\end{figure*}

Motivated by the above questions, in this paper, we study the implementation of JC-Hubbard model with superconducting circuits including weakly anharmonic SQs. We consider the lowest three levels for the SQs because the high level anharmonicity becomes very large. The influence of this anharmonicity on the ground-state property is discussed, and it shows that the Mott insulator to superfluid quantum phase transition still can be obtained even when the SQs have weak anharmonicity. Interestingly, the SQs becomes an equal spaced three level system when the anharmonicity approaching zero. This ultimately leads to the result that on-site two-photon repulsion effect disappear corresponding to the ground-state phase diagram without single excitation Mott lobe. This work also offers a method to manipulate the Mott lobes of ground state by adjusting the anharmonicity of SQs. It is useful for exploring the quantum criticality of the lattice of superconducting circuits with weak anharmonicity.

This paper is organized as follows: In Sec. II, we introduce the model considered here, i.e., a two-dimensional lattice including coupled microwave cavities and weakly anharmonic SQs.
In Sec. III, we discuss the ground-state property of this lattice and show the phase transition between Mott-insulator and superfluid phase. The complete ground-state phase diagram of our model is presented.
Conclusions are given in Sec. IV.

\section{Our model}
We consider an expanded JC-Hubbard model depicted in Fig.\,\ref{Fig1}, including a on-site interaction between the SQs with three levels and the microwave resonator.
Here the third level $|f\rangle$ of SQs is introduced due to the weak anharmonicity of the SQs (such as the transmon qubits). The system Hamiltonian is
\begin{align}
H=\sum_i H_i-H_{\rm hop}-\mu\sum_i N_i,
\label{model}
\end{align}
where the on-site expanded JC interaction is described by~\cite{Wu1996,Wu1997}
\begin{align}
 H_{i}=&\omega_c a_i^\dag a_i +\omega_q|e\rangle_i\langle e|+(2\omega_q+\Delta)|f\rangle_i\langle f|\nonumber\\
    &+(\lambda_1 a_i^\dag|g\rangle_i\langle e| + \lambda_2 a_i^\dag|e\rangle_i\langle f|+h.c.).
    \label{JC}
\end{align}
Here $a$ and $a^\dag$ are the annihilation and creation operators of the cavity model, and $\omega_c$, $\omega_q$ are the frequencies of the cavity and SQs. We denote the qubit-cavity frequency detuning $\delta=\omega_q-\omega_c$. The qubit-cavity coupling strengths are denoted by $\lambda_1$, $\lambda_2$, and in general $\lambda_2=\sqrt{2}\lambda_1=\sqrt{2}\lambda$. Note that, here we only consider three levels of SQs, i.e., $|g\rangle$, $|e\rangle$ and $|f\rangle$, since the high level anharmonicity are much larger than $\Delta$. As shown in the following discussion, this expanded JC interaction also will offer the photon-photon repulsion except for the case of $\Delta=0$.

Hamiltonian $H_{\rm hop}$ [i.e., the second term of Eq.\,(\ref{model})] denotes the photon hopping between the nearest-neighbor cavities with hopping rate $J$ and $H_{\rm hop}=-J\sum_{\langle i,j\rangle}(a^{\dagger}_i a_j+a^{\dagger}_j a_i)$. The chemical-potential term of Eq.\,(\ref{model}) is obtained by treating our system within the grand canonical ensemble. Here the on-site chemical potential is $\mu$ and $N_i=a_i^\dag a_i+|e\rangle\langle e|+2|f\rangle\langle f|$ is the conserved number of polaritons. Note that, in contrast to the
situation in ultracold-atom systems, the chemical potential $\mu$ can not be obtained directly in the present coupled system. However, this problem can be solved by devising appropriate preparation
schemes to access states with different
mean polariton numbers~\cite{Angelakis2007}.

The competition between the photon repulsion and photon hopping effects induces the occurrence of Mott insulator to superfluid phase transition. To obtain insight properties of the total system, we apply the mean field approximation~\cite{Oosten2001,Oosten2003,Milburn2000} into Hamiltonian (\ref{model}), i.e., $a_i^\dag a_j=\langle a_i^\dag\rangle a_j+\langle a_j \rangle a_i^\dag-\langle a_i^\dag \rangle\langle a_j \rangle$, and introduce a superfluid order parameter $\psi=\langle a_i \rangle$, which is taken to be real~\cite{Oosten2001}. Then the system Hamiltonian becomes
\begin{align}\label{meanf}
H_{\rm MF}=&\sum_i\{H_i-zJ\psi(a_i^\dag+a_i)+zJ|\psi|^2\nonumber
\\
&-\mu(a_i^\dag a_i+|e\rangle_i\langle e|+2|f\rangle_i\langle f|)\},
\end{align}
where $z=3$ is the number of nearest neighbours, which ensures the validity of mean field approximation used in our model. In principle, our model also could be implemented in the one-dimensional array of superconducting circuit, where the mean field approximation can not be used. The quantum phase transition in the case of one-dimensional array might be discussed in our following works.

\section{Mott insulator to superfluid phase transition}
\subsection{On-site photon-photon repulsion}
To show the on-site photon-photon repulsion in the present system, we first discuss the eigenvalues and eigenstates of system in the weak hopping limit, i.e., $J=0$. Here the conserved number of polaritons $N_i$ decides that the present system has the photon dependent eigenvalue. The system ground state is $|g,0\rangle$ corresponding to $N_i=0$. When $N_i=1$, the Hamiltonian (\ref{JC}) is equivalent to a JC model, and hence its eigenvalues and eigenstates are
\begin{subequations}
\begin{align}\label{JC-e}
E_{|\pm,1\rangle}&=\omega_c+\frac{\delta}{2}\pm\sqrt{\lambda^2+\delta^2/4},
\\
|\pm,1\rangle&=\frac{(\delta/2\pm\sqrt{\lambda^2+\delta^2/4})|g,1\rangle+\lambda|e,0\rangle}{\sqrt{2\lambda^2+\delta^2/2\mp\sqrt{\lambda^2+\delta^2/4}\delta}}.
\end{align}
\end{subequations}
When $N_i\geq2$, in the subspace {$|g,n\rangle$, $|e,n-1\rangle$, $|f,n-2\rangle$}, the on-site Hamiltonian $H_{i}$ can be written as
\begin{equation}\label{matrix}
H_{i}=
\begin{pmatrix}
  n\omega_c & \sqrt{n}\lambda & 0\\
  \sqrt{n}\lambda & n\omega_c+\delta & \sqrt{2(n-1)}\lambda \\
  0 & \sqrt{2(n-1)}\lambda & n\omega_c+2\delta+\Delta
\end{pmatrix}.
\end{equation}

In Fig.\,\ref{Fig2}, we present some of eigenvalues of the system, and they split into three branches denoted by $|+,n\rangle$ (the upper branch), $|0,n\rangle$ (the middle branch), and $|-,n\rangle$ (the lower branch).
The on-site repulsion effects increasing with anharmonicity $|\Delta|$ is shown in this eigenspectrum. Moreover, the eigenvalues of system in the single excitation subspace is constant with respective to $\Delta$, and an single excitation shift $\eta$ from the harmonic level structure is obtained. This ultimately leads to the result that two photon repulsion disappears when $\Delta=0$, where the two excitation shift from the harmonic level structure is $2\eta$ [see Figs.\ref{Fig2}(b,d,e)]. In this case, the dressed states structure of system from the ground state to the second excitation state becomes harmonic, as shown in Fig.\,\ref{Fig1}(d).
As shown in the following discussion, the disappearance of two photon repulsion leads to the disappearance of single excitation Mott lobe for certain chemical potential $\mu$.

\subsection{Ground state phase diagram}
Besides the above on-site photon repulsion interaction, our system also has the on-site chemical potential term and the photon hopping between the neighboring cavities, as shown in total Hamiltonian (\ref{model}).
Normally, the competition between the photon repulsion and photon hopping induces the occurrence of quantum phase transition. Specifically, the system is in the Mott insulator phase when the photon repulsion dominates over the photon hopping effect. Now the system has a fixed number of excitations per site. On the contrary, the system is in the superfluid phase.
This phase transition could be characterized qualitatively by the superfluid order parameter $\psi$, i.e., $\psi=0$ and $\psi\neq0$ corresponding to the Mott insulator and superfluid phases, respectively.

In the limit of weak photon hopping $J/\lambda\ll1$, the boundaries of the Mott lobes can be decided by associating Eqs.\,(\ref{JC}) and (\ref{meanf}).
To obtain the ground state of system, we only need to consider the negative branch due to $E_{|-,n\rangle}<E_{|+,n\rangle}$ and $E_{|-,n\rangle}<E_{|0,n\rangle}$. The total number of excitations per site changes at $E_{|-,n+1\rangle}-\mu(n+1)=E_{|-,n\rangle}-\mu n$, which decides the boundaries of Mott lobes. In Fig.\,\ref{Fig3}, we present some of boundaries of Mott lobes in the weak limit of $J$. Interestingly, the boundary between the zero-excitation and single-excitation Mott lobes is flat with respective to the anharmonicity $\Delta$. This flat boundary cross the boundary between the single excitation and two excitation Mott lobes at $\Delta=0$, which divides into two regions in Fig.\,\ref{Fig3}. Before this cross point, i.e., $\Delta<0$, the zero and one excitation lobes are covered, and they only appear after this cross point, i.e., $\Delta>0$. When $\Delta=0$, only the single excitation Mott lobe is covered. Physically, this is because the two excitation nonlinearity disappears when $\Delta=0$ in our model, as shown in Fig.\,\ref{Fig2}. This result is consistent with the previous discussions in subsection A.

By diagonalizing the Hamiltonian (\ref{meanf}), in Fig.\,\ref{Fig4} we plot the complete phase diagram of the mean-field solution under the condition of qubit-cavity resonance, i.e., $\delta=0$.
Similar to the normal JC Hubbard model, rich dynamics is illustrated in Fig.\,\ref{Fig4}. The superfluid phase corresponds to the regions where $\psi\neq0$, and the stable ground state of at each site is a coherent state. The Mott insulator phase corresponds to the case of $\psi=0$, and its number of excitations increases with $\mu$. Interestingly, it shows that the anharmonicity $\Delta$ of SQ will influence the Mott insulator lobes effectively. The Mott lobe with one excitation disappears at $\Delta=0$, and the largest size of Mott lobes is found when $\Delta>0$. This result is consistent with Fig.\,\ref{Fig3}. In a short summary, the weak anharmonicity of SQ will not destroy the Mott lobes completely, but it will influence its size and make special Mott lobe disappear.

To confirm the excitation numbers in each Mott lobes, we could calculate the average number of excitations per site in the grand canonical ensemble $\rho$ given by
\begin{equation}\label{excitation}
  \rho=-\frac{\partial E_{g(\psi=\psi_{min})}}{\partial\mu}.
\end{equation}
In Fig.\ref{Fig5}, we plot $\rho$ as a function of $J/\lambda$, $(\mu-\omega)/\lambda$ for different anharmonicity $\Delta$ under the qubit-field resonant condition $\delta=0$. These stair-step shapes indicate the regions of Mott lobes with fixed excitation numbers. It also clearly show that the single excitation lobe disappears when $\Delta=0$ due to the two excitation nonlinearity disappearing. The maximum Mott lobes with fixed excitation number appears when $\Delta>0$.

\section{Conclusion}
In conclusion, we have studied the quantum phase transition in an extended JC-Hubbard model, where the on-site qubit has weak anharmonicity and an auxiliary level is introduced.
We showed that this weak anharmonicity will not destroy the occurrence of Mott insulator to superfluid quantum phase transition. It will influence the size of Mott lobes via changing the on-site photon-photon repulsion.
The single excitation Mott lobe is covered when the anharmonicity $\Delta$ disappear. Our results show that the quantum phase transition from Mott insulator to superfluid phase still could be implemented even in the superconducting circuits including weak anharmonic SQs. Moreover, we have shown that the anharmonicity also could be used to manipulate the ground state excitation number on per site in the coupled lattice of superconducting circuits.
\\
\\
\\
\begin{acknowledgments}
This work is supported by the National Key Research and Development Program of China grant
2016YFA0301203-02, the National Science Foundation of China (Grant Nos. 11374116, 11574104 and 11375067).
\end{acknowledgments}

\end{document}